\documentstyle[12pt, psfig]{article}

\newcommand{\NP}{{\em Nucl.\ Phys.\ }}
\newcommand{\PL}{{\em Phys.\ Lett.\ }}
\newcommand{\PR}{{\em Phys.\ Rev.\ }}
\newcommand{\PRP}{{\em Phys.\ Rep.\ }}

\newcommand{\MPL}{{\em Mod.\ Phys.\ Lett.\ }}
\newcommand{\PRL}{{\em Phys.\ Rev.\ Lett.\ }}
\newcommand{\IJMP}{{\em Int.\ J.\ Mod.\ Phys.\ }}

\newcommand{\tr}{{\rm Tr}}
\newcommand{\re}{{\rm Re}}
\newcommand{\im}{{\rm Im}}

\newcommand{\inn}{\!\cdot\!}

\begin{document}
\pagestyle{plain}
\setcounter{page}{1}

\baselineskip16pt

\begin{titlepage}

\begin{flushright}
PUPT-1642\\
hep-th/9608127\\
\end{flushright}
\vspace{20 mm}

\begin{center}
{\huge Dynamics of Dirichlet-Neumann Open~Strings on D-branes
}

\vspace{5mm}

\end{center}

\vspace{10 mm}

\begin{center}
{\large Akikazu Hashimoto}

\vspace{3mm}

Joseph Henry Laboratories\\
Princeton University\\
Princeton, New Jersey 08544

\end{center}

\vspace{2cm}

\begin{abstract}
\noindent Method for computing scattering amplitudes of open strings
with Dirichlet boundary on one end and Neumann boundary condition on
the other is described.  Vertex operator for these states are
constructed using twist fields which have been studied previously in
the context of Ashkin-Teller model and strings on orbifolds.  Using
these vertex operators, we compute the three- and four-point
scattering amplitudes for (5,9) strings on 5-branes and 9-branes.  In
the low energy limit, these amplitudes are found to be in exact
agreement with the field theory amplitudes for supersymmetric
Yang-Mills coupled to hypermultiplets in 6-dimensions. We also
consider the 1-brane 5-brane system and compute the amplitude for a 
pair of (1,5) strings to collide and to escape the brane as a closed
string.  (1,5) strings are found to be remarkably selective in their
coupling to massless closed strings in NS-NS sector; they only couple
to the dilaton.
\end{abstract}

\noindent

\vspace{2cm}
\begin{flushleft}
August 1996
\end{flushleft}
\end{titlepage}
\newpage

\renewcommand{\baselinestretch}{1.1}  
\addtolength{\baselineskip}{0.5ex}
\section{Introduction}
\label{Intro}

String solitons and non-perturbative dynamics of string theory have
been a subject of fascination for some time.  Remarkable progress has
been made recently in the understanding of this subject following the
realization that certain solitons in string theory admit explicit
description in terms of D-branes \cite{joeReview}. These solitons are
charged with respect to the Ramond-Ramond fields \cite{polchinski} and
have been identified as the source of various non-perturbative
phenomena\footnote{For recent review and references, see
\cite{joeReview} and references therein.}.

One of the useful properties of D-branes is the fact that they admit a
perturbative description at weak coupling in terms of the open strings
attached to them.  These open strings can interact amongst themselves,
or interact with closed string degrees of freedom outside.  Their
dynamics can be described using traditional string perturbation theory
with slight modifications. Techniques for studying this kind of
physics were developed recently in \cite{KT,GHKM,gm,decay}. These
studies revealed string scale as the dynamical length scale governing
the physics of D-brane fluctuations\footnote{There is however, growing
evidence for sub-string length scales arising from D-brane
interactions \cite{Bachas,Shenker,DanFerrSund,PouliotKabat,DKPS96}.}
and rich dynamics on the D-brane world volume.

In this article, we extend the results of \cite{KT,GHKM,gm,decay} to
open strings which have Neumann boundary condition on one end and
Dirichlet boundary condition on the other end (the ND strings).  These
strings can exist when several D-branes of differing dimensionality,
say, a 1-brane and a 5-brane, are simultaneously present.  There are
reasons to be interested in the perturbative dynamics of ND strings.
For instance, bound states of 1-branes and 5-branes have been well
studied in the context of black hole entropy counting
\cite{cm,ghas,HMS96}. When both 1-branes and 5-branes are abundant, it
is the (1,5) strings which dominates entropically, and their dynamics
is of interest in attempts to continue weak coupling calculations to
the strong coupling region \cite{DasMathur96-3}.

The need for calculational techniques for ND strings becomes even more
acute when one attempts to study emission and absorption cross
sections, as was done in \cite{DMW96,DasMathur96-2,SteveIgor}.  For simple
D-brane configurations, one can rely on the Dirac-Born-Infeld action
\cite{DBI} to teach us how the world volume degrees of freedom on the
D-brane couple to the space-time fields.  For complicated D-brane
bound states, such as the ones involving multiple 1-branes and
5-branes, no analogous description of the effective coupling is known.
We are therefore forced to perform string theory calculations along
the lines of \cite{decay} to deduce the effective coupling and to
attempt to reconstruct the generalized Dirac-Born-Infeld action.  In
this article, we will develop the techniques necessary to carry out this
program.

The first step in the perturbative study of string amplitudes is the
construction of vertex operators.  When an ND vertex is inserted,
boundary conditions on the world sheet change.  The ND vertex operator
therefore necessarily contains boundary condition changing
operators. We will show that {\em twist fields}, familiar from
previous studies in the Ashkin-Teller model \cite{ashkin-teller},
off-shell string dynamics \cite{bershad86}, and orbifolds
\cite{orbifold} play the necessary role.  We will use these twist
fields to construct a vertex operator in conformal gauge which
satisfies all the requirements \cite{FMS}. We will find that this
vertex operator closely resembles the vertex operator for twisted
states in $Z_2$ orbifolds of type II theories \cite{orbifold}.

In the remainder of the article, we will use these vertex operators to
compute tree-level scattering amplitudes.  We will follow the general
strategy of \cite{decay} and relate the physics of open string sector
alone and that of open/closed interactions.  For concreteness, we will
focus on type IIB theories.  The 1-brane 5-brane system will always be
in the back of our mind as a concrete example.  In discussing open
string sector, however, we will consider the 5-brane 9-brane system
instead, as on-shell polarizations for gauge particles are absent on
the 1-brane world volume.  They are related simply by $T$-duality
and/or dimensional reduction.

A comment is in order regarding the choice of vacuum.  We will work in
a background where all the fields have vanishing vacuum expectation
value.  This is the simplest background from the point of view of
worldsheet conformal field theory.  (For more general backgrounds, we
must consider non-linear sigma models.) This is a special point in the
moduli-space of vacua where branes are bound at threshold. They are
free to fluctuate in shape and size \cite{doug,witinst}.

Let us also note in passing that techniques developed in this article
are applicable to a wider variety of D-brane configurations. For
example, the 1-brane 5-brane configuration featured in this article is
$T$-dual to a configuration of intersecting D-branes
\cite{seninter}. Intersecting branes have been found to play an important
role in the D-brane construction of 4-dimensional black holes
\cite{IgorTseytlin,VijayFinn}. We hope to say more about open string
dynamics on intersecting branes in the future.

The organization of this article is as follows.  In section 2, we
study world sheets with mixed Neumann and Dirichlet boundary
conditions, and construct the vertex operator for ND strings.  In
section 3, we compute the three and four-point amplitude for these ND
strings and compare with low energy effective theory.  In section 4,
we compute the Hawking emission amplitude for colliding (1,5) and
(5,1) strings.  We conclude in section 5.

\section{World-sheets with Dirichlet and Neumann boundaries}

\begin{figure}
\centerline{\psfig{figure=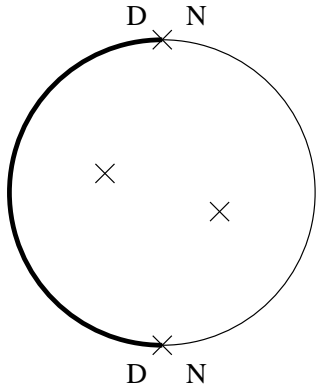}}
\caption{Open string world sheet with two ND vertex operators
inserted.  The boundary condition of the disk changes as one crosses
the ND vertices along the boundary.}
\label{figa}
\end{figure}

The goal of this section is to construct the vertex operator for the
ND strings.  Because these states are open string states, the vertex
operators are inserted along the boundary of the world sheet.  The
fact that these states are ND strings means that the boundary
condition of the world sheet changes from Dirichlet to Neumann and
vice versa at the insertion point of its vertex operator (see figure
\ref{figa}).
The ND open string vertex operator must necessarily contain a boundary
condition changing operator.  To learn how one constructs such an
operator, it is useful to study the scalar Green's function on a disk
with mixed Dirichlet-Neumann boundary conditions.  By bringing the
charges near the point where the boundary condition changes, we can
infer the singularity structure of the operator product expansion of
the twist operator and the scalar field.

Calculation of scalar Green's function is a simple exercise in
electrostatics.  It is simplest to map the disk onto a strip
parameterized by complex variable $\{ \zeta: -\infty < \re\ \zeta <
+\infty, 0 < \im\ \zeta < \pi\}$. The strip can be mapped onto
half-plane by $z = e^{\zeta}$, and the half plane can be mapped on to
the disk as usual.  We impose Neumann boundary condition at $\im\
\zeta = 0$ and Dirichlet boundary condition at $\im\ \zeta=\pi$.  On
the half-plane, this corresponds to Dirichlet boundary condition on
the negative real-axis and Neumann boundary condition on the positive
real axis.  The change of boundary condition occurs at $z=0$ and
$z=\pm \infty$.

We can determine the potential at $\zeta$ on a strip due to a charge
at $\omega$ by placing image charges of charge $(-1)^n$ at
$(\omega+2\pi n\, i)$ and $(\bar{\omega}+2 \pi n\, i)$. These image
charges ensure that appropriate boundary conditions are satisfied.
The potential at $\zeta$ is therefore given by
\begin{eqnarray}
G(\zeta,\bar{\zeta};\omega,\bar{\omega}) & = & 
\left(\sum_n (-1)^n \left\{ \ln \left[\zeta - (\omega+2\pi n\, i)\right] 
+\ln \left[\zeta - (\bar{\omega}+2\pi n\, i)\right] \rule{0in}{2ex} \right\} \right)
+ (\zeta \leftrightarrow \bar{\zeta}) \nonumber \\
& = & \left\{\ln \left[ 
\frac{\prod_{n\ {\rm even}} (\zeta-\omega+2 \pi n\, i)}
    {\prod_{n\ {\rm odd}} (\zeta-\omega+2 \pi n\, i)} \right] 
+ (\omega \leftrightarrow \bar{\omega})  \right\}
+ (\zeta \leftrightarrow \bar{\zeta}) .
\end{eqnarray}
The infinite product in the above expression can be evaluated
explicitly in terms of matching expression with appropriate residues
and poles. Mapping to half plane using $z = e^\zeta, w = e^\omega$, we
find
$$
G(z,w) = \frac{\prod_{n\ {\rm even}} (\zeta-\omega+2 \pi n\, i)}
    {\prod_{n\ {\rm odd}} (\zeta-\omega+2 \pi n\, i)}
= \ln \left[
\frac{1-e^{\frac{\zeta-\omega}{2}}}{1+e^{\frac{\zeta-\omega}{2}}}
      \right]
= \ln\left[
	\frac{1-\sqrt{\frac{z}{w}}}{1+\sqrt{\frac{z}{w}}}\right].
$$
This expression corresponds to scalar potential on a half plane where
boundary condition changing operator is inserted at $z=0$ and $z=\pm
\infty$.  We might as well conformally deform the boundary condition
changing points to arbitrary points along the boundary of the half plane.
This is straightforward to do, and we find
$$
G(z_3,z_4) = \ln \left[
\frac{1-\sqrt{\frac{z_{13}z_{24}}{z_{14}z_{23}}}}
     {1+\sqrt{\frac{z_{13}z_{24}}{z_{14}z_{23}}}}
\right]
$$
where the boundary condition changing points are $z_1$ and $z_2$,
while $z_3$ and $z_4$ are the arguments of the Green's functions.  The
electrostatic Green's function is essentially the two-point correlation
function of a scalar boson.  However, in practice, it is the electric
field, not the electrostatic potential, which makes a good conformal
field. We are therefore interested in the correlation function of the
primary field $\partial X$.
\begin{equation}
\frac{\langle \sigma(z_1) \sigma(z_2) \partial X(z_3) \partial X(z_4) \rangle}
     {\langle \sigma(z_1) \sigma(z_2) \rangle}
= \partial_{z_3} \partial_{z_4} G(z_3,z_4)
\label{green}
\end{equation}
Here we have introduced a yet to be identified boundary condition
changing operator $\sigma(z)$ inserted at $z_1$ and $z_2$. Now let us
probe the singularity structure of operator product expansion of
$\partial X$ and $\sigma$ by studying the $z_3 \rightarrow z_1$ limit.
By explicit calculation, we find that
$$
\partial_{z_3} \partial_{z_4} G(z_3,z_4) = \frac{1}{{z_{13}}^{1/2}}
F(z_1,z_2,z_4) + \ldots
$$
where $F$ is some function. What this suggests is that we should look
for an operator algebra which includes $\partial X$ and $\sigma$ such
that their leading operator product expansion is of the form
$$
\partial X(z) \sigma(0) = \frac{1}{z^{1/2}} \tau + \ldots
$$
where $\tau$ is another operator with dimension $1/2$ greater than the
dimension of $\sigma$. It just so happens that a closed algebra
consisting of $\partial X$, $\sigma$, and $\tau$ alone exists as a
conformal limit of Ashkin-Teller model \cite{ashkin-teller,bershad86}.
These fields have dimensions
$$
[\sigma]= \frac{1}{16};\qquad
[\tau]=\frac{9}{16}
$$
and their operator algebra is given by
\begin{eqnarray}
\partial X^\alpha(z) \sigma^\beta(w) & = & \delta^{\alpha \beta} 
(z-w)^{-1/2} \tau^\beta(w) + {\rm Reg} \nonumber \\
\partial X^\alpha(z) \tau^\beta(w) & = & \frac{1}{2}
\delta^{\alpha \beta} (z-w)^{-3/2} \sigma^\beta(w) + 2 \delta^{\alpha \beta}
(z-w)^{-1/2} \partial_w \sigma^\beta(w) + {\rm Reg} \nonumber.
\end{eqnarray}
Fields $\sigma$ and $\tau$ are referred to as ``twist fields'' and
``excited twist fields,'' respectively. These fields have also been
used in the vertex operators for type II twisted states on a $Z_2$
orbifold \cite{orbifold}. In the context of orbifolds, the square root
branch cut in the operator product expansion implied that as $\partial
X$ circumscribes $\sigma$, it picks up a minus sign, implying that $X$
and $-X$ are identified (i.e. orbifolded).  In open strings, there is
no such thing as a twisted state. Because the branch point is inserted
at the boundary of the world sheet, we can maintain single valued
operator algebra on the world sheet without any space-time
identifications by simply pushing the branch cut to the outside the
world sheet.  It is therefore not necessary to orbifold the spacetime.
Instead, the square-root branch cut exchanges the boundary condition
between Dirichlet and Neumann.  This can also be seen from the mode
expansion of twisted states
$$
X(z) = \sum_{r \in Z+\frac{1}{2}} a_r z^r.
$$
which is identical to the mode expansion of ND strings.

Correlation functions involving these twist fields have been studied
in \cite{ashkin-teller,bershad86,orbifold} using conformal field
theory techniques.  They are in agreement with equation (\ref{green})
computed using the electrostatic method.  By now, it is clear that
these twist fields implement the change in boundary conditions.

Having found the operator responsible for changing the boundary
condition, let us proceed to construct the vertex operators for ND
string states. To be concrete, we will focus on Neveu-Schwarz states on
5-branes and 9-branes, oriented as follows
\begin{equation}
\begin{tabular}{ccccccccccc}
        &0&1&2&3&4&5&6&7&8&9\\
5-brane:&N&N&N&N&N&N&D&D&D&D \\
9-brane:&N&N&N&N&N&N&N&N&N&N \\
\end{tabular}
\label{9brane-orient}
\end{equation}
which breaks $SO(1,9)$ down to $SO(1,5) \times SO(4)$. At times, we
will refer to ND strings as (5,9) strings to indicate the brane on
which the strings end. In order to implement the appropriate boundary
condition on the world sheet, we need to insert a boundary condition
changing operator for directions $\{6789\}$.  Supersymmetry requires
that we also change the boundary condition for the world sheet
fermions.  In the $\{6789\}$ directions, Neveu-Schwarz states have
Ramond boundary conditions. Therefore, the vertex operator for the ND
state must also contain spin fields for $\{6789\}$ directions. This is
consistent with the oscillator analysis \cite{joeReview} where the
spectrum was shown to contain a massless state which is singlet under
$SO(1,5)$ and a chiral spinor of definite chirality under $SO(4)$.
The simplest natural candidate for the ND vertex operator then is
\begin{equation}
V(z)  = \lambda_{iJ} \,
u_a e^{-\phi} {S^\alpha} \Delta e^{i\, 2pX}(z).
\label{vertex}
\end{equation}
The factor of $\lambda_{iJ}$ describes the Chan-Paton degrees of
freedom which label the brane on which the strings end. The indices
$i$ and $J$ transform under fundamental and anti-fundamental
representations of $U(Q_5)$ and $U(Q_9)$ respectively.  The $SO(4)$
polarization of this state is given by $u_a$. The spin field $S(z)$ is
given by the usual bosonization rule, except it only involves the
$\{6789\}$ directions
$$
S(z) = e^{\pm i\, \phi_{67}/2 \pm i\, \phi_{89}/2}
$$
and has dimension $1/4$.  $\Delta$ is the boundary condition changing
operators for the directions $\{6789\}$:
$$
\Delta(z) = \sigma^6 \sigma^7 \sigma^8 \sigma^9 (z)
$$
and has dimensions $1/16 \times 4 = 1/4$. 

The charge conjugate of (5,9) string is a (9,5) string. Charge
conjugation flips the index structure of the Chan-Paton factor and
changes the chirality of the spinor.  The vertex operator then has 
the structure
\begin{equation}
V(z)  = \lambda_{Ji} \,
u_\beta e^{-\phi} {S_\beta} \Delta e^{i\, 2pX}(z).
\end{equation}

The superghost contribution to the conformal weight was fixed by the
requirement that the total vertex operator be of conformal weight one.
Not surprisingly, an identical expression was derived in the context
of orbifolds using superfield techniques and picture changing
\cite{orbifold}. Furthermore, it is amusing to note that we can
reproduce the formula for the zero-point energy \cite{joeReview}
$$
-\frac{1}{2}+ \frac{1}{8} \nu
$$
by simply requiring the vertex operator to have conformal weight one. Here,
$\nu$ is the number of ND dimensions. The $-1/2$ comes from the
dimension of ghosts, and $1/8$ comes from conformal weight per
embedding dimension of $1/16$ for the twist field and $1/16$ for the
spin field.

We have found an explicit expression (\ref{vertex}) for the vertex
operator of ND states. This is the main result of this article. In
principle, one can compute any amplitudes involving ND states along
this line.  In the following sections, we will work out a few examples.

\section{Open string dynamics of ND strings}

In this section, we will study the tree-level dynamics of ND open
strings.  All external states in the amplitudes we consider in this
section will be open strings.  The low energy dynamics of these open
strings are given by $N=1$ super Yang-Mills with gauge fields $U(Q_5)$
and $U(Q_9)$ dimensionally reduced to six dimensions, and coupled to a
hypermultiplet in fundamental and anti-fundamental representations of
$U(Q_5)$ and $U(Q_9)$, respectively.  Their interactions are mostly
determined by supersymmetry and gauge invariance.  The bosonic part of
the action is given by \cite{doug2,juanthesis}
\begin{eqnarray}
S &=&  \int d^6 x\
\frac{1}{4}\tr (F_{\mu \nu} F^{\mu \nu})
+\frac{1}{2}\tr [ ( \partial_\mu A_M + g [A_\mu,A_M])^2] \nonumber \\
&&\qquad
 +\frac{1}{4} \tr (F'_{\mu \nu} F'^{\mu \nu})
+\frac{1}{2} \tr [ ( \partial_\mu A'_M + g' [A'_\mu,A'_M])^2] \nonumber \\
&& \qquad\qquad
+\left|(\partial_\mu + g A_\mu^{a} T^a + g' A_{\mu}'^a T^a ) \chi\right|^2
+ \frac{1}{4} g^2
\sum_{aMN}{D_{MN}^{a}}^2 + 
\frac{1}{4} {g'}^2\sum_{aMN}{{D'}_{MN}^{a}}^2.
\label{action}
\end{eqnarray}
Gauge fields $A$ and $A'$ transform under $U(Q_5)$ and $U(Q_9)$
respectively. Indices $\mu$ and $\nu$ run from 0 to 5. Indices $M$ and
$N$ runs from 6 to 9. Fields $A_M$ arise from dimensional reduction of
gauge fields and are scalars.  The hypermultiplet is described by
$\chi$.  In addition to gauge coupling, hypermultiplets couple to the
gauge field by $D$-terms in the action.  They are given by
\begin{eqnarray}
{D_{MN}^a}^2 &=&  \left( f_{bc}^a A_M^b A_N^c
+ \chi^{\dagger} T^a \Gamma_{MN} \chi \right) \nonumber \\
{{D'}_{MN}^a}^2 &=&  \left( {f'}_{bc}^a {A'}_M^b {A'}_N^c
+ \chi^{\dagger} {T'}^a \Gamma_{MN} \chi \right) . \nonumber
\end{eqnarray}
In the remainder of this section we will compute some tree level
amplitudes involving the ND strings and verify that they are consistent
with (\ref{action}) in the low energy limit.

\subsection{Three point functions}

One simplest kind of amplitudes one can compute in string theory are
the three point tree amplitudes.  Let us first focus on these.  A
moment's thought should convince the reader that an even number of
boundary condition changing operators is necessary along the
boundary. Indeed, terms cubic in $\chi$'s are absent in the action
(\ref{action}).  On the other hand, there is nothing inconsistent with
a coupling of $A_\mu$ or $A_M$ with two $\chi$'s from the world sheet
point of view.  Let us consider each case separately.

\subsubsection{$\chi \chi A_\mu$ coupling}

The relevant vertex operators are
\begin{eqnarray}
V_1 & = & \lambda_1^{iJ}\,
 {u_1}_\alpha\,
 e^{-\phi} \Delta S^\alpha e^{i\, 2k_1 X}(z_1) \nonumber \\
V_2 & = & \lambda_2^{Jk}\,
 {u_2}^\beta e^{-\phi}\,
 \Delta S_\beta e^{i\, 2k_2 X}(z_1) \nonumber \\
V_3 & = & \Lambda_3^{ki}\,
 {\xi_3}_\mu\,
( \partial X^\mu + (i\, 2 k_3 \inn \psi) \psi^\mu)
e^{i\, 2 k_3 X}(z_3).
\end{eqnarray}
The total superghost charge is $-2$. The amplitude is given by
\begin{eqnarray}
A &=& \int \frac{dz_1\, dz_2\, dz_3}{V_{CKG}}
\langle V_1(z_1) V_2(z_2) V_3 (z_3)\rangle  \nonumber \\
&=& \tr (\lambda_1 \lambda_2 \Lambda_3)
(z_{12} z_{13} z_{23}) \cdot \langle e^{-\phi}(z_1) e^{-\phi}(z_2)\rangle 
\cdot {\xi_3}_\mu \langle 
e^{i\, 2 k_1 X}(z_1)
e^{i\, 2 k_2 X}(z_2) \partial X^\mu
e^{i\, 2 k_3 X}(z_3) \rangle \nonumber \\
&& \qquad\qquad \cdot {u_1}_\alpha {u_2}^\beta \langle S^\alpha(z_1) S_\beta(z_2)\rangle
\cdot \langle \Delta(z_1) \Delta(z_2) \rangle.
\end{eqnarray}
All of the correlation functions are standard except possibly for
$$
\langle \Delta(z_1) \Delta(z_2) \rangle = \frac{1}{z_{12}^{1/2}}
$$
which follows from the fact that $\Delta$ has conformal weight $1/4$.
The amplitude evaluates to
$$
A = \tr(\lambda_1 \lambda_2 \Lambda_3)\,
{u_1} _\alpha {u_2}^\alpha\, {\xi_3}_\mu  (k_1^\mu - k_2^\mu)
$$
which is in agreement with the coupling due to the gauge interaction term
$$
S =\left| (\partial_\mu+ g A_\mu) \chi \right|^2
$$
in the action (\ref{action}).

\subsubsection{$\chi \chi A_M$ coupling}

The vertex operators are the same as in the $\xi \xi A_\mu$ case, but
the polarization vector $\xi_M$ will now point in the $\{6789\}$
direction.  Just as in the previous case, the amplitude is given by
\begin{eqnarray}
A &=& \int \frac{dz_1\, dz_2\, dz_3}{V_{CKG}}
\langle V_1(z_1) V_2(z_2) V_3 (z_3)\rangle  \nonumber \\
&=& \tr (\lambda_1 \lambda_2 \Lambda_3)
(z_{12} z_{13} z_{23}) \cdot \langle e^{-\phi}(z_1) e^{-\phi}(z_2)\rangle 
\cdot \langle 
e^{i\, 2 k_1 X}(z_1)
e^{i\, 2 k_2 X}(z_2) 
e^{i\, 2 k_3 X}(z_3) \rangle \nonumber \\
&& \qquad\qquad \cdot {u_1}_\alpha {u_2}^\beta \langle S^\alpha(z_1) S_\beta(z_2)\rangle
\cdot {\xi_3}_M \langle \Delta(z_1) \Delta(z_2) \partial X^M(z_3) \rangle.
\end{eqnarray}
Here, we encounter a new correlation function
$$ \langle \Delta(z_1) \Delta(z_2) \partial X^M (z_3) \rangle. $$
In the electrostatic language, however, this amounts to measuring the
electric field in the presence of boundary condition changing
operators.  Since no sources of electric fields are present, this
correlation function must vanish.  This implies that $\chi \chi A_M$
three point function also vanishes.  This is in agreement with the
absence of $\chi \chi A_M$ coupling term in the low-energy effective
action (\ref{action}).

\subsection{Four point functions}

The next simplest thing to the three-point functions are the
four-point functions. These amplitudes are often written in terms of
relativistic invariants
$$s = 4 k_1 \inn k_2 = 4 k_3 \inn k_4,\qquad
t = 4 k_1 \inn k_4 = 4 k_2 \inn k_3, \qquad
u = 4 k_1 \inn k_3 = 4 k_2 \inn k_4.
$$
Generically, these four point amplitudes exhibit Regge-pole behavior,
indicating exchange of a tower of massive intermediate string states.
The low energy limit is gotten by expanding the amplitude to leading
order in derivatives.  In this approximation, the amplitude
gets contributions only from the leading $s$ and $t$-channel pole, as
well as contact terms in the low energy effective theory.

As was noted in the previous section, consistency requires that there
be even number of external $\chi$'s in a given process.  Amplitudes
containing exactly two $\chi$'s are particularly simple. There are
three such amplitudes: $\chi \chi A_\mu A_\nu$, $\chi \chi A_M A_N$,
and $\chi \chi A_\mu A_M$. Let us consider each case separately.

\subsubsection{$\chi \chi A_\mu A_\nu$ coupling}
\label{sectionA}

The relevant vertex operators are
\begin{eqnarray}
V_1 & = & \lambda_1^{iJ}\,
 {u_1}_\alpha \,
 e^{-\phi} \Delta S^\alpha e^{i\, 2k_1 X}(z_1) \nonumber \\
V_2 & = & \lambda_2^{Jk}\,
 {u_2}^\beta e^{-\phi}\,
 \Delta S_\beta e^{i\, 2k_2 X}(z_1) \nonumber \\
V_3 & = & \Lambda_3^{kl}\,
 {\xi_3}_\mu\,
( \partial X^\mu + (i\, 2 k_3 \inn \psi) \psi^\mu)
e^{i\, 2 k_3 X}(z_3) \nonumber \\
V_4 & = & \Lambda_4^{li}\,
 {\xi_4}_\nu\,
( \partial X^\nu + (i\, 2 k_4 \inn \psi) \psi^\nu)
e^{i\, 2 k_4 X}(z_4).
\end{eqnarray}
The total superghost charge is $-2$. The amplitude is given by
$$
A = \int \frac{dz_1\, dz_2\, dz_3\, dz_4}{V_{CKG}}
\langle V_1(z_1) V_2(z_2) V_3 (z_3) V_4(z_4)\rangle .
$$
Although generally speaking one must compute these correlation
functions in a usual manner, in this particular case there is a
short-cut. The correlation function factorizes into parts involving
world sheet fields with space-time index $\{012345\}$ and parts
involving space-time index $\{6789\}$.  Similar factorization takes
place in a computation of $A_M A_N A_\mu A_\nu$.  We can therefore cut
the $\{6789\}$ part of the $\chi \chi A_\mu A_\nu$ correlation
function and paste it into the more familiar $A_M A_N A_\mu A_\nu$
amplitude which takes the form
$$
A = \frac{\Gamma(s)\Gamma(t)}{\Gamma(1+s+t)}
K(\xi_1,k_1;\xi_2,k_2;\xi_3,k_3;\xi_4,k_4).
$$
The kinematic factor $K$ is the familiar open string kinematic factor
\cite{schwarzPhysRep} except for the fact that in our dimensionally
reduced system, $\xi_1$ and $\xi_2$ are orthogonal to $\xi_3$,
$\xi_4$, and all the $k_i$'s
$$
K(\xi_1,k_1;\xi_2,k_2;\xi_3,k_3;\xi_4,k_4)
= \left( 
   {{t\,u\,}\over 4} \xi_3\inn \xi_4+
  u\,k_1\inn \xi_4\,k_2\inn \xi_3 + 
   t\,k_1\inn \xi_3\,k_2\inn \xi_4 \right)\, \xi_1 \inn \xi_2.
$$
Upon comparison of $\{6789\}$ part of the correlation function, we
note that the only difference between $\chi\chi A_\mu A_\nu$ and $A_M
A_N A_\mu A_\nu$ is that $\xi_1 \inn \xi_2$ is replaced by
$u_1  u_2$. Therefore, the amplitude of interest is given by
$$
A = \sum_{3\leftrightarrow 4}
 \frac{\Gamma(s)\Gamma(t)}{\Gamma(1+s+t)}
 \left( 
   {{t\,u\,}\over 4} \xi_3\inn \xi_4+
  u\,k_1\inn \xi_4\,k_2\inn \xi_3 + 
   t\,k_1\inn \xi_3\,k_2\inn \xi_4 \right)\, 
u_1 u_2 \ \tr(\lambda_1 \lambda_2 \Lambda_3 \Lambda_4).
$$
In the low energy limit, this reduces to
\begin{eqnarray}
A&=&
-\frac{1}{8} \xi_3 \inn \xi_4\, \tr(\lambda_1 \lambda_2 \{\Lambda_3,\Lambda_4\})
\nonumber \\
&&\frac{1}{s} \left[\frac{1}{8} (u-t) \xi_3 \inn \xi_4
- \xi_3 \inn k_2\, \xi_4 \inn k_1 
+ \xi_3 \inn k_1\, \xi_4 \inn k_2 \right] 
\tr(\lambda_1 \lambda_2 [\Lambda_3, \Lambda_4])\nonumber \\
&& - \frac{1}{t} \xi_3 \inn k_2\, \xi_4 \inn k_1 
\tr(\lambda_1 \lambda_2 \Lambda_3 \Lambda_4) 
- \frac{1}{u} \xi_3 \inn k_1 \, \xi_4 \inn k_2
\tr(\lambda_1 \lambda_2 \Lambda_4 \Lambda_3) .
\end{eqnarray}

All terms in the above expression also arise from the low-energy
effective action.  For example, 
$$\frac{1}{8} \xi_3 \inn \xi_4\, \tr(\lambda_1 \lambda_2 \{\Lambda_3,\Lambda_4\})
$$
is due to the contact term in the gauge coupling
$$
\left| (\partial_\mu + g A_\mu^a T_a) \chi\right|^2,
$$
and the term proportional to $1/s$ is due to exchange of massless
gluons between $\chi\chi A$ and $A^3$ three-point vertices.

\subsubsection{$\chi \chi A_M A_N$ coupling}

The vertex operator is the same as in the $\chi \chi A_\mu A_\nu$ case
except for the the fact that polarization vectors ${\xi_3}_M$ and
${\xi_4}_N$ point in the $\{6789\}$ direction. The correlation function
of interest takes the form
\begin{eqnarray}
&&\langle V_1(z_1) V_2(z_2) V_3(z_3) V_4(z_4) \rangle
= \langle e^{-\phi}(z_1) e^{-\phi}(z_2) \rangle \cdot
\langle e^{i\, 2 k_1 X} (z_1)
e^{i\, 2 k_2 X} (z_2)
e^{i\, 2 k_3 X} (z_3)
e^{i\, 2 k_4 X} (z_4)
\rangle \nonumber \\
&& \qquad\qquad\qquad\left\{\rule{0ex}{2ex}  4 {k_3}^\mu {k_4}^\nu \langle  \psi_\mu(z_3)  \psi_\nu(z_4)
 \rangle 
\cdot  \langle S^\alpha(z_1) S_\beta(z_2) \psi^M(z_3) \psi^N(z_4) \rangle 
\cdot \langle \Delta(z_3) \Delta(z_4) \rangle \right. \nonumber \\
&& \qquad\qquad\qquad\qquad
\left. + \langle S^\alpha(z_1) S_\beta(z_2) \rangle \cdot 
\langle \Delta(z_1) \Delta(z_2) \partial X (z_3) \partial X (z_4) \rangle
\rule{0ex}{2ex}\right\}.
\end{eqnarray}
Twist field correlation function $\langle \Delta(z_1) \Delta(z_2)
\partial X(z_3) \partial X(z_4) \rangle$ can be deduced from
(\ref{green}). The spin field correlation function
\begin{eqnarray}
\langle S^\alpha(z_1) S_\beta(z_2) \psi^M(z_3) \psi^N(z_4)\rangle
&=& (z_{14} z_{24} z_{13} z_{23})^{-1/2}(z_{12})^{-1/2} (z_{34})^{-1} 
\nonumber\\
&&\qquad \times
\frac{1}{2}
\left\{ \delta^{MN} \delta^\alpha_\beta (z_{14} z_{23}+z_{13} z_{24})
- [\Gamma^{M N}]^\alpha_\beta \,z_{12} z_{34} \right\} \nonumber
\end{eqnarray}
was worked out in \cite{KLLSW87}.  Fixing conformal Killing volume by
fixing the location of vertex operators at $\{ -z, z, 1, -1\}$, the
amplitude reduces to a simple integral expression
\begin{equation}
A = \int dz 
\left[\frac{4z}{(1+z)^2}\right]^s \left[\frac{(1-z)^2}{(1+z)^2}\right]^t
\left[(s-1) \frac{1+z^2}{z^2} \delta^\alpha_\beta g^{MN} -
\frac{2s}{z} [\Gamma^{MN}]^\alpha_\beta\right]
\tr(\lambda_1 \lambda_2 \Lambda_3 \Lambda_4)
\label{chichiAMAN}.
\end{equation}
Performing the integral, one obtains
$$A = \sum_{3\leftrightarrow 4}
\left((s+2t)\, \xi_3 \inn \xi_4\,  u_1 u_2\, -
 s \, (u_1\Gamma^{MN}u_2)\, {\xi_3}_M {\xi_4}_N \right) 
\frac{\Gamma(s)\Gamma(\frac{1}{2}+t)}{\Gamma(\frac{1}{2}+s+t)}
\tr(\lambda_1 \lambda_2 \Lambda_3 \Lambda_4).
$$
The half integer poles in the $t$-channel may seem unusual. These
states arise from modes excited in the direction for which the
world-sheet boundary condition is ND, as the oscillator number of
these excitations is half integral. In the low energy limit, this
amplitude reduces to
$$
A =\frac{t-u}{s} \xi_3 \inn \xi_4\, u_1 u_2 \tr(\lambda_1 \lambda_2
[\Lambda_3, \Lambda_4]) 
+ (u_1 \Gamma^{MN} u_2) {\xi_3}_M {\xi_4}_N \tr(\lambda_1 \lambda_2
[\Lambda_3, \Lambda_4]) .
$$
From the low energy effective field theory point of view, the first
term corresponds to exchange of gluon between $\chi \chi A$ and $A^3$
three-point vertices, and the second term corresponds to contact term
arising from the $D$-term.  The low energy limit of $\chi\chi A_M A_N$
amplitude is found to be consistent with the low-energy effective field
theory.

\subsubsection{$\chi \chi A_\mu A_M$ coupling}

The vertex operators are the same as in the previous two subsections,
except for the orientation of the polarization vectors.  Upon inspection,
one finds that the correlation function takes the form
\begin{eqnarray}
\lefteqn{\langle V_1(z_1) V_2(z_2) V_3(z_3) V_4(z_4) \rangle
=}\nonumber \\
&& \langle \Delta(z_1) \Delta(z_2) \partial X^M(z_4) \rangle \cdot
\langle S^\alpha(z_1) S_\beta(z_2) \rangle \times \left\{ {\rm
ghosts\ and\ 012345\hbox{-}field\ correlator} \right\} \nonumber \\
&& + \langle \Delta(z_1) \Delta(z_2) \rangle \cdot
\langle S^\alpha(z_1) S_\beta(z_2) \psi^M(z_4) \rangle \times \left\{ {\rm
ghosts\ and\ 012345\hbox{-}field\ correlator} \right\}. \nonumber \\
\end{eqnarray}
Since $ \langle \Delta(z_1) \Delta(z_2) \partial X^M(z_4) \rangle$ and
$\langle S^\alpha(z_1) S_\beta(z_2) \psi^M(z_4) \rangle$ both vanish,
it follows that $\chi \chi A_\mu A_M$ four-point amplitude also
vanishes.  This too is consistent with the low-energy effective theory, as
neither $\chi\chi A_M$ and $A_\mu A_\nu A_M$ three point coupling nor
$\chi\chi A_\mu A_N$ contact term is present in the theory.

\section{Hawking radiation from decay of ND strings}

So far we have focused on open string sector and have found agreement
between the dynamics of ND states in string theory and the low energy
effective field theory.  These ND strings can also interact with the
closed strings propagating outside the brane through the usual
mechanism of open-closed string interaction.  An example of such an
interaction is a process where a pair of open string collides and
escapes the brane by becoming a closed string.  This kind of process
mimics Hawking radiation if these D-branes are truly black holes
\cite{cm} and has recently been studied more carefully in
\cite{DMW96,DasMathur96-2,SteveIgor}.

The near extremal black hole studied in \cite{cm,ghas,HMS96} is a
system of 1-branes and 5-branes.  This is related to the system of
5-branes and 9-branes considered in the previous section by
T-dualizing along $\{2345\}$ directions.  For reference, we list
the orientation of these branes below:
\begin{equation}
\begin{tabular}{rcccccccccc}
                 &0&1&2&3&4&5&6&7&8&9\\
         1-brane:&N&N&D&D&D&D&D&D&D&D \\
         5-brane:&N&N&D&D&D&D&N&N&N&N \\
\end{tabular}.
\label{15brane-orient}
\end{equation}
Directions $\{6789\}$ are compactified, and the 5-brane wraps around
the internal space and becomes a string parallel to the D1-brane.  The
compactification breaks the Lorentz group from $SO(1,9)$ to $SO(1,5)
\times SO(4)_I$. The string further breaks $SO(1,5)\times SO(4)_I$
down to $SO(1,1) \times SO(4)_E \times SO(4)_I$.

We further compactify the $\hat{1}$ direction.  By taking this
compactification radius larger than the compactification radius in
$\{6789\}$ directions, the 1-brane 5-brane system describes a
macroscopic black string wrapped around the long direction which was
discussed in \cite{ghas}. On the other hand, if the $\hat{1}$ radius
is small, this system describes a black hole in 5-dimensions.

When the numbers of 1-branes and 5-branes are both large, strings
stretching between 1-branes and 5-branes dominate entropically.
These are precisely the ND strings we have been studying.  It would be
quite interesting to compute the Hawking radiation rate for the decay
of (1,5) strings.

Fortunately, it was shown recently in \cite{decay} how amplitudes of
this type can be computed quite effortlessly in terms of open-string
four-point functions.  These are precisely the four-point amplitudes
we have been considering in the previous section.  In the remainder of
this section, we will apply the technique of \cite{decay} to compute
the decay amplitude of (1,5) strings.

We will only compute the decay into massless closed string states in
the Neveu-Schwarz sector.  These states are gravitons, antisymmetric
tensors, and the dilaton.  In the context of black hole (string)
entropy counting \cite{cm,ghas,HMS96}, these fields are dimensionally
reduced to $5+1$ dimension by compactifying along the $\{6789\}$
directions.  It is therefore convenient to consider polarizations
which transform differently under $SO(1,5)\times SO(4)$
separately. The polarization breaks down under $SO(1,5)\times SO(4)$
into the following groups: $\varepsilon_{\mu \nu}$, $\varepsilon_{\mu
N}$, $\varepsilon_{MN}$. Polarization tensor $\varepsilon_{\mu \nu}$
describes the graviton, the antisymmetric tensor, and the dilaton in
$5+1$ dimensions.  States with polarization $\varepsilon_{\mu N}$
describes states which transform as a vector under $SO(5,1)$ Lorentz
group.  Finally, $\varepsilon_{MN}$ is a scalar under $SO(1,5)$.  Note
that this grouping is precisely analogous to cases $\chi \chi A_\mu
A_\nu$, $\chi \chi A_\mu A_N$, and $\chi \chi A_M A_N$ which we
considered separately in the previous section.

What makes these amplitudes so easy to compute is that under
suitable prescription of fixing the conformal killing group, the
Hawking amplitude and the open string 4-point amplitude take on an
identical form.  The only thing which distinguishes between these
amplitudes is the contour of integration.  We refer readers to
\cite{decay} for the details concerning this trick.

We denote the momenta of incoming (1,5) and (5,1) strings by $p_1$ and
$p_2$.  Because these strings are attached to the 1-brane, momentum
vectors are constrained to lie in $\{01\}$ plane.  We denote the
momentum of the closed string by $q^\mu$.  There is one kinematic
invariant for this process given by
$$
t = 2 p_1 \inn q = 2 p_2 \inn q = -2 p_1 \inn p_2.
$$
It will be convenient to restrict $q^\mu$ to lie in $\{012345\}$
space, as the open string correlation functions in section 3 were
computed for this case.  Since $\{6789\}$ directions are compactified,
this corresponds to considering only states neutral with respect to
the Kaluza-Klein gauge fields in the context of Hawking radiation
calculation.

Let us start with the $\varepsilon_{\mu\nu}$ case.  In the
corresponding $\chi \chi A_\mu A_\nu$ calculation, the correlation
function was found to be identical in form to that of $A^4$ amplitude.
Without doing any work, it follows that the $\varepsilon_{\mu\nu}$
amplitude must be of the form identical to what was found in
\cite{decay}, namely
$$
A = \frac{\Gamma(-2t)}{\Gamma(1-t)^2} K(1,2,3)
$$
where $K(1,2,3)$ is obtained from the kinematic factor we found in
section \ref{sectionA} by imposing the kinematic constraint $-2s=t=u$
and making the substitution $\varepsilon \inn D \rightarrow \xi_3
\otimes \xi_4$.  In the low-energy limit, this amplitude reduces to
\begin{equation}
A =   u_1 u_2 \left(
	  p_2\inn \varepsilon \inn p_1 
	+ p_1 \inn \varepsilon \inn p_2
   	+ {{t}\over 4}\, \tr \, \left(D \inn \varepsilon \right) \right)
\label{lowenergy}
\end{equation}
where the trace runs only over $\{012345\}$ directions.  To enumerate
the physical closed string states involved in this process, it is
convenient to express the amplitude explicitly in terms allowed
momentum and polarization vectors.  Without loss of generality, we can
set
$$ q^\mu = \{q_0, q_1, \vec{q} \}$$
where $\vec{q}$ points in some $\{2345\}$ direction. In six
dimensions, there are four physical polarizations transverse to
$q^\mu$:
$$
\begin{array}{rcrcr}
\epsilon_\perp&=&\{0,&\frac{|\vec{q}|}{q_0},&-\frac{q_1}{q_0} \hat{q}\} \\
\epsilon_A     & = & \{0,&0,&\vec{\epsilon}_A\ \} \\
\epsilon_B     & = & \{0,&0,&\vec{\epsilon}_B\ \} \\
\epsilon_C     & = & \{0,&0,&\vec{\epsilon}_C\ \} \\
\end{array}
$$
where $\{\hat{q},\vec{\epsilon}_A,\vec{\epsilon}_B,\vec{\epsilon}_C\}$
are mutually orthogonal set of vectors in the $\{2345\}$ plane.  The
most general polarization for which the amplitude (\ref{lowenergy}) is
non-vanishing is of the form
\begin{equation}
\varepsilon^{\mu \nu} =
C_\perp \epsilon_\perp^\mu \epsilon_\perp^\nu + 
C_A \,\epsilon_A^\mu \epsilon_A^\nu +
C_B \,\epsilon_B^\mu \epsilon_B^\nu +
C_C \,\epsilon_C^\mu \epsilon_C^\nu
\label{generalpolarization}
\end{equation}
where $C_\perp$, $C_A$, $C_B$, and $C_C$ are numerical coefficients.
Substituting (\ref{generalpolarization}) back into (\ref{lowenergy}),
we arrive at
$$
A = -\frac{t}{4} \left( C_\perp + C_A + C_B + C_C \right).
$$
The only state for which this amplitude is non-vanishing is the
dilaton, suggesting a low-energy effective coupling of the form
\begin{equation}
S =\phi \partial^\mu \chi^\alpha \partial_\mu \chi_\alpha.
\label{coupling}
\end{equation}
Here, indices $\mu$ and $\nu$ run over the dimensions of the 1-brane
world volume $\{01\}$. 

For the case of $\varepsilon_{\mu N}$, we found that the correlation
function vanished in the corresponding open string 4-point amplitude.
This means that a pair of (1,5) strings can not decay into an
$\varepsilon_{\mu N}$ state.

Finally, let us consider the $\varepsilon_{MN}$ case. The integral
expression for the corresponding open-string amplitude was written
down in (\ref{chichiAMAN}).  The prescription of \cite{decay} is to
make a change of variables $z = ix$ and integrate over $-\infty < x<
\infty$.  When this is done (\ref{chichiAMAN}) becomes
$$
A = \int_{-\infty}^\infty dx\, \left[\frac{(1+x^2)^2}{16x^2}\right]^t
\left( (2t+1)\frac{(1-x^2)}{x^2} \delta^\alpha_\beta g^{MN}
- \frac{2s}{ix} [\Gamma^{MN}]^\alpha_\beta\right).
$$
This integral vanishes as first term is odd under $x \rightarrow 1/x$
and the second term is odd under $x \rightarrow -x$.  (1,5) strings
can not decay into $\varepsilon_{MN}$ state.

To summarize, the only massless NS-NS state to which a pair of (1,5)
string can collide and decay into is the dilaton.  No other fields
seem to couple to the (1,5) strings.

\section{Conclusions}

In this article we studied the perturbative dynamics of open strings
in the ND sector.  By studying the singularity in electrostatic
potential near the boundary condition changing point on a half-plane,
we were led to twist fields as a necessary ingredient.  Using these
twist fields we constructed the vertex operators satisfying the usual
requirements and carrying appropriate spacetime quantum numbers. The
resulting vertex operator turned out to closely resemble the vertex
operator for type II twisted states on $Z_2$ orbifolds.  In rough
terms, ND states are the open string analogue of twisted
states\footnote{Twisted sector of open strings was considered recently
in \cite{Blum96} from a different point of view.}.

Using these vertex operators, we computed several string amplitudes.
The effective low-energy dynamics on the world volume was shown to be
in perfect agreement with the string theory calculation.  

We then followed \cite{decay} and computed the leading amplitude for a
pair of (1,5) strings to collide and escape the brane.  These
amplitudes exhibit the same Regge-pole structure previously found for
the DD strings \cite{decay}. Somewhat surprisingly, (1,5) strings are
found to be quite selective in their choice of decay-particles.  They
appear to couple only to the dilaton.

The tools developed in this article provides us with control over
perturbative dynamics of ND strings. It would be interesting to
explore the consequence of string theory dynamics to black hole
physics.

\section*{Acknowledgments}

We are grateful to David Gross for valuable discussions, and Igor
Klebanov for discussions and critical reading of the manuscript.  We
also thank Juan Maldacena, Emil Martinec, Joe Polchinski, Arvind
Rajaraman, Sanjaye Ramgoolam, and Steve Shenker for illuminating
discussions.  This work was supported in part by DOE grant
DE-FG02-91ER40671, the NSF Presidential Young Investigator Award
PHY-9157482, and the James S. McDonnell Foundation grant No.  91-48.

\addtolength{\baselineskip}{-0.5ex}


\begin{thebibliography}{10}

\bibitem{joeReview}
J. Polchinski, S. Chaudhuri, and C. V. Johnson, ``Notes on D-Branes,''
  NSF-ITP-96-003, {\tt hep-th/9602052}.

\bibitem{polchinski}
J. Polchinski, \PRL {\bf 75} (1995) 4724. 

\bibitem{KT}
I. R. Klebanov and L. Thorlacius, \PL {\bf B371} (1996) 51-56.

\bibitem{GHKM}
S.S.\ Gubser, A.\ Hashimoto, I.R.\ Klebanov, and J.M.\ Maldacena,
  ``Gravitational lensing by $p$-branes,'' {\tt hep-th/9601057}.

\bibitem{gm}
M.R.\ Garousi and R.C.\ Myers, ``Superstring Scattering from D-Branes,'' {\tt
  hep-th/9603194}.

\bibitem{decay}
A. Hashimoto and I.R. Klebanov, ``Decay of Excited D-branes,'' {\tt
  hep-th/9604065}.

\bibitem{Bachas}
C. Bachas, ``D-Brane Dynamics,'' {\tt hep-th/9511043}.

\bibitem{Shenker}
S. H. Shenker, ``Another Length Scale in String Theory?'' Rutgers preprint
  RU-95-53, {\tt hep-th/9509132}.

\bibitem{DanFerrSund}
U. Danielsson, G. Ferretti, and B. Sundborg, Stockholm preprint USITP-96-03,
  {\tt hep-th/9603081}.

\bibitem{PouliotKabat}
D. Kabat and P. Pouliot, ``A Comment on Zero-Brane Quantum Mechanics,'' Rutgers
  preprint RU-96-17, {\tt hep-th/9603127}.

\bibitem{DKPS96}
M.R. Douglas, D. Kabat, P. Pouliot, and S. Shenker, ``D-Branes and Short
  Distances in String Theory,'' {\tt hep-th/9608024}.

\bibitem{cm}
C. Callan and J. Maldacena, ``D-brane Approach to Black Hole Quantum
  Mechanics,'' Princeton preprint PUPT-1591, {\tt hep-th/9602043}.

\bibitem{ghas}
G. Horowitz and A. Strominger, ``Counting States of Near Extremal Black
  Holes,'' {\tt hep-th/9602051}.

\bibitem{HMS96}
G. Horowitz, J. Maldacena, and A. Strominger, ``Nonextremal Black Hole
  Microstates and U-duality,'' {\tt hep-th/9603109}.

\bibitem{DasMathur96-3}
S.R. Das and S.D. Mathur, ``Interactions involving D-branes,'' {\tt hep-th
  9607149}.

\bibitem{DMW96}
A. Dhar, G. Mandal, and S.R. Wadia, ``Absorption v.s. Decay of Black Holes in
  String Theory and T-Symmetry'' TIFR-TH-96-26, {\tt hep-th/9605234}.

\bibitem{DasMathur96-2}
S.R. Das and S.D. Mathur, ``Comparing Decay Rates for Black Holes and
  D-Branes,'' MIT-CTP-2546, {\tt hep-th/9606185}.

\bibitem{SteveIgor}
S.S. Gubser and I.R. Klebanov, ``Emission of Charged Particles from Four- and
  Five-dimensional Black Holes,'' {\tt hep-th/9608108}.

\bibitem{DBI}
R. G. Leigh, \MPL {\bf A4} (1989) 2767.

\bibitem{ashkin-teller}
A.B. Zamolodchikov, \NP {\bf B285} (1987) 481-503.

\bibitem{bershad86}
M. Bershadski, \IJMP {\bf A1} (1986) 443-449.

\bibitem{orbifold}
L. Dixon, D. Friedan, E. Martinec, and S. Shenker, \NP {\bf B282} (1987) 13-73;
  S. Hamidi and C. Vafa, \NP {\bf B279} (1987) 465.

\bibitem{FMS}
D.~Friedan, E.~Martinec, and S.~Shenker, \NP {\bf B271} (1986) 93.

\bibitem{doug}
M. Douglas, ``Branes within Branes,'' {\tt hep-th/9512077}.

\bibitem{witinst}
E. Witten, \NP {\bf B460} (1996) 541.

\bibitem{seninter}
A. Sen, \PR {\bf D53} (1996) 2874-2894 

\bibitem{IgorTseytlin}
I.R. Klebanov and A.A. Tseytlin, ``Intersecting M-branes as Four-Dimensional
  Black Holes,'' PUPT-1616, Apr 1996. {\tt hep-th/9604166}.

\bibitem{VijayFinn}
V. Balasubramanian, F. Larsen, ``On D-branes and Black Holes in
  Four-Dimensions,'' PUPT-1617, Apr 1996. {\tt hep-th/9604189}.

\bibitem{doug2}
M.R. Douglas, ``Gauge Fields and D-Branes,'' RU-96-24, {\tt hep-th/9604198}.

\bibitem{juanthesis}
J.M. Maldacena, ``Black Holes in String Theory,'' PhD thesis, Princeton
  University, June 1996 {\tt hep-th/9607235}.

\bibitem{schwarzPhysRep}
J.H.\ Schwarz, \PRP {\bf 89} (1982) 223.

\bibitem{KLLSW87}
V.A. Kostelecky, O. Lechtenfeld, W. Lerche, S. Samuel, and S. Watamura, \NP
  {\bf B288} (1987) 173-232.

\bibitem{Blum96}
J.D. Blum, ``F Theory Orientifolds, M Theory Orientifolds, and Twisted
  Strings,'' {\tt hep-th/9608053}.

\end{thebibliography}
\end{document}